\UseRawInputEncoding
\documentclass[12pt,a4paper,final]{iopart}

\usepackage{iopams}  
\usepackage{graphicx}
\usepackage[breaklinks=true,colorlinks=true,linkcolor=blue,urlcolor=blue,citecolor=blue]{hyperref}
\usepackage{caption}
\usepackage{subcaption}

\begin{document}

\title[Project ThaiPASS]{Project ThaiPASS: International Outreach Blending Astronomy and Python}

\author{James D. Keegans, Richard J. Stancliffe, Lawrence E. Bilton, Claire R. Cashmore, Brad K. Gibson, Mikkel T. Kristensen, Thomas V. Lawson, Marco Pignatari, Iraj Vaezzadeh}
\address{E.A. Milne Centre for Astrophysics, The University of Hull, Cottingham Road, Kingston-Upon-Hull, HU6 7RX, United Kingdom}

\ead{J.Keegans-2016@hull.ac.uk}

\author{Benoit C\^{o}t\'{e},}
\address{Konkoly Observatory, Research Centre for Astronomy and Earth Sciences, MTA Centre for Excellence, Budapest H-1121, Hungary}
\address{ELTE E\"otv\"os Lor\'and University, Institute of Physics, Budapest 1117, Hungary}

\author{and Siri Chongchitnan}
\address{Warwick Mathematics Institute, University of Warwick, Coventry, CV4 7AL, United Kingdom.}

\begin{abstract}
We present our outreach program, the \textit{Thailand-UK Python+Astronomy Summer School} (ThaiPASS), a collaborative project comprising UK and Thai institutions and assess its impact and possible application to schools in the United Kingdom. Since its inception in 2018, the annual ThaiPASS have trained around 60 Thai high-school students in basic data handling skills using Python in the context of various astronomy topics, using current research from the teaching team. Our impact assessment of the 5 day summer schools show an overwhelmingly positive response from students in both years, with over 80\% of students scoring the activities above average in all activities but one. We use this data to suggest possible future improvements. We also discuss how ThaiPASS may inspire further outreach and engagement activities within the UK and beyond. 
\end{abstract}

\vspace{2pc}


\section{Introduction}

\textit{ThaiPASS} (Thailand-UK Python+Astronomy Summer School) is an overseas development assistance (ODA) outreach project focusing on building students' programming skills for the future. The project was spearheaded by the University of Hull in collaboration with the National Astronomy Research Institute of Thailand, and has since garnered additional participating institutions. \textit{ThaiPASS} brings UK professional astronomers to Thailand, to train 16-18 year-old Thai students in using Python for data handling, analysis and mining. Student engagement was fostered through the use of astronomy and astrophysics as the bulk of the teaching matter. Astronomy and astrophysics have been shown, through initiatives such as Galaxy Zoo \cite{raddick2009galaxy}, to capture the interest of a broad range of individuals. Using astronomy, we also highlight the importance of data science for the modern economy and STEM subjects. The first {\it ThaiPASS} was launched in 2018 under the auspices of the STFC (Science and Technology Facilities Council) Newton Fund scheme, running annually, with funding now secured through 2022. {\it ThaiPASS} activities have now been extended for teacher Career Professional Development (CPD) training events in the UK.

In this report, we discuss findings from the first two events, held in 2018 and 2019, in Chiang Mai, Thailand \footnote{{\it ThaiPASS} 2020 has been postponed to 2021, due to the Covid19 pandemic.}. In total, 60 students and 19 teachers from 30 Thai schools participated in the first two {\it ThaiPASS} events. Teachers participated in the activities with the goal of applying this experience in the classroom, increasing the effective impact of the summer school to include students at each of these 30 schools. Each summer school comprised a series of lectures and hands-on sessions over the course of a week, focusing on various topics in astronomy. Students brought their own laptops, and additional online computing facilities were provided in collaboration with the University of Victoria, Canada, as part of their Cyberhubs initiative \cite{herwig2018cyberhubs}. All data and data mining tools used in the summer school were produced by our lecturing team and their collaborators, much of which is currently published \cite{pignatari2016nugrid,ritter2018nugrid,ritter2016nupycee}.

This paper is structured as follows. We first discuss the motivation for {\it ThaiPASS} in Section \ref{motivation}. We then present the astronomy and computing material covered at the summer schools in Section \ref{content}. In Section~\ref{analysis}, we discuss the feedback from participants, and the challenges faced. Finally, we outline our plans for future events in Section \ref{plans}. A full overview of {\it ThaiPASS} can also be found at the respective 2018 and 2019 summer school websites\footnote{\href{https://indico.narit.or.th/event/81}{https://indico.narit.or.th/event/81} and \href{https://indico.narit.or.th/event/120}{https://indico.narit.or.th/event/120}, respectively.}.

\section{Motivation}\label{motivation}

\subsection{Why ThaiPASS?}
Project {\it ThaiPASS} is a human capacity building project with a strong ODA resonance: ICT and data handling capabilities are two of the key technical skills young people need to further their professional development and obtain the highly skilled jobs which drive the modern economy.

One of the most effective ways to achieve the goals of ODA is to strengthen Thailand’s knowledge-based economy. ThaiPASS addresses this by directly reaching out to young people in their final years of secondary education. These students may still be undecided about university and career choices, and they may not have had any experience with programming and data-handling. By attending ThaiPASS, they will be equipped with these skills and STEM knowledge that may not be readily available to them otherwise. We  accomplish all this by using STEM frontier research disciplines like astronomy and computational astrophysics, which many school students find to be an extremely exciting and inspirational subject.

From an ODA perspective, the combination of ICT education and astronomy presents a highly cost-effective and high-impact way to further economic development, since STEM knowledge and skills are transferable to many different sectors and underpins economic advances in health systems, education and infrastructure. The kind of training offered at ThaiPASS aligns with and supports existing initiatives  concerned with the development of a knowledge-based high-tech economy which Thailand envisages itself to be.

The motivations for ThaiPASS also resonate with the United Nation's \textit{Sustainable Development Goals} (SDG) to end poverty, protect the planet, and ensure prosperity for all. Project ThaiPASS addresses the following SDGs.

\begin{itemize}
    \item \textit{SDG4: Ensuring inclusive and quality education for all and promoting lifelong learning}. ThaiPASS brings Thai school students and teachers in contact with international researchers, who will train them in scientific skills that they may not have access to (or afford) otherwise. The training received at ThaiPASS will be current and relevant, and will be a seed for lifelong learning of STEM subjects.
    Our target students are those who have a passion and aptitude in science, maths and English from Thai schools across the country (those in highly exclusive, international schools are not targeted). We  made concerted effort in targeting schools that are in underprivileged areas, and  strongly urged schools to nominate both male and female participants. We note that the gender ratio was 50:50 for ThaiPASS'18.

    \item \textit{SDG8: Promoting inclusive and sustainable economic growth, employment and decent work for all}. A young person with ICT, data-handling and visualisation skills is a highly employable person within the STEM sector and beyond. A STEM-enabled generation of young people is essential for the future of Thailand’s information-based economy.

    \item \textit{SDG9: Build resilient infrastructure, promote sustainable industrialization and foster innovation.} ThaiPASS represents an investment in human capacity building, which is crucial to achieving sustainable development, empowering communities, and generating a high sustainable income with a low environmental impact.
\end{itemize}
\subsection{Why Python?}

The Python programming language is one of the most popular programming languages in astronomy and the wider sciences. Its position as an open source, high level, simple-to-use programming language has made it a popular choice in many areas of research. Although other programming languages (such as C, FORTRAN) are still used for intensive computational tasks, data science needs are widely carried out using Python. Its popularity and widespread use in science research make it invaluable as a tool for any student who wishes to pursue a career in astronomy. Python is also highly desirable in industry, and as such also provides a useful learning opportunity for students wishing to pursue careers outside of academia. 

Python, being a high-level programming language, often uses natural language for built-in functions, making it more approachable that some other languages. The use of  Jupyter notebooks also improves accessibility, as the interactive environment allows the students to easily test and edit code with a simple GUI. Python also has a huge array of modules available which are invaluable in scientific computing. These include modules such as:

\begin{enumerate}
    \item Astropy - packages containing astronomy related functions
    \item Matplotlib - plotting functionality
    \item NumPy - mathematical operations not native to Python
    \item SciPy - contains functions, statistical tools, numerical solvers for differential equations and many other features
\end{enumerate}

All of the above factors make Python programming a highly desirable and useful skill for students, particularly those planning to go into a career in STEM.


\section{Content of ThaiPASS}
\label{content}

All of the materials covered at  ThaiPASS'18 and 19 are available at the E. A. Milne Centre's GitHub webpages\footnote{ \href{https://github.com/Milne-Centre/ThaiPASS2018}{https://github.com/Milne-Centre/ThaiPASS2018} and \href{https://github.com/Milne-Centre/ThaiPASS2019}{https://github.com/Milne-Centre/ThaiPASS2019} respectively.}. These materials comprise Jupyter notebooks and data files, which formed the basis of activities at ThaiPASS, as well as basic introductory tutorials consisting of scaffolded notebooks with a variety of simple tasks. As the most current version of much of the material, the authors suggest using the 2019 repository where possible (a description of the contents of each repository is provided in their respective read-me). We present a summary of some of the materials here.

\subsection{Astronomy content}
\label{astro_topics}

Astronomy has been a compulsory part of secondary education in Thailand since 2001, and a surprisingly wide range of topics are covered in Thai schools\footnote{Detail of astronomy education in Thailand (and many other countries) can be found at the IAU Office of Astronomy for Education \href{https://www.haus-der-astronomie.de/oae/worldwide}{https://www.haus-der-astronomie.de/oae/worldwide}}. Nevertheless, there are disparities in the depth and quality of astronomy education, notably amongst non-private schools, mainly due to the lack of teachers with specialist training in physics and astronomy, and inadequate  provision for after-school astronomy activities. At ThaiPASS, we made a decision to ensure that as much of the astronomy content is as self-contained and accessible as possible to 16-18 year olds. We also included introductory discussions of modern research topics outside the standard syllabus. To date, the topics covered include:
\begin{itemize}
    \item  \textit{nucleosynthesis}, the production of elements and isotopes in stars;
    \item  \textit{galactic chemical evolution} (GCE), the process by which material synthesised by stars is mixed into the wider galaxy;
    \item \textit{galaxies and dark matter}, which discusses how the rotation curves of galaxies lead to the idea that there is unseen matter in the Universe;
    \item \textit{supernova}, describing how stars explode and how ejecta material spreads out into space.
\end{itemize} 
During morning lectures, students were given an introduction to the main ideas and concepts of each topic, and in the afternoon they were given tasks to complete using data from recent publications (See for example \cite{pignatari2016nugrid} and \cite{cote2018origin}). 

We note that the lectures on stellar evolution build on and extend topics found in the Scottish  \href{https://www.sqa.org.uk/files_ccc/AHPhysicsCourseSpec.pdf}{Advanced Higher Specifications}, the AQA  \href{https://filestore.aqa.org.uk/resources/physics/specifications/AQA-7407-7408-SP-2015.PDF}{A-level syllabus} and other exam boards in the UK (with some intersection with the Thai syllabus). Common elements in these specifications concern the competition between two opposing forces: the burning of nuclear fuel in a star, and its collapse under gravity.  The material covered here allows for a deeper understanding of the processes and pathways involved in the production of new nuclei in stars. This topic is expanded further in the second set of lectures on nucleosynthesis described below.

\subsubsection{Nucleosynthesis}
\label{nucleosynthesis_sec}

\begin{figure}
    \centering
    \includegraphics[width=0.7\textwidth]{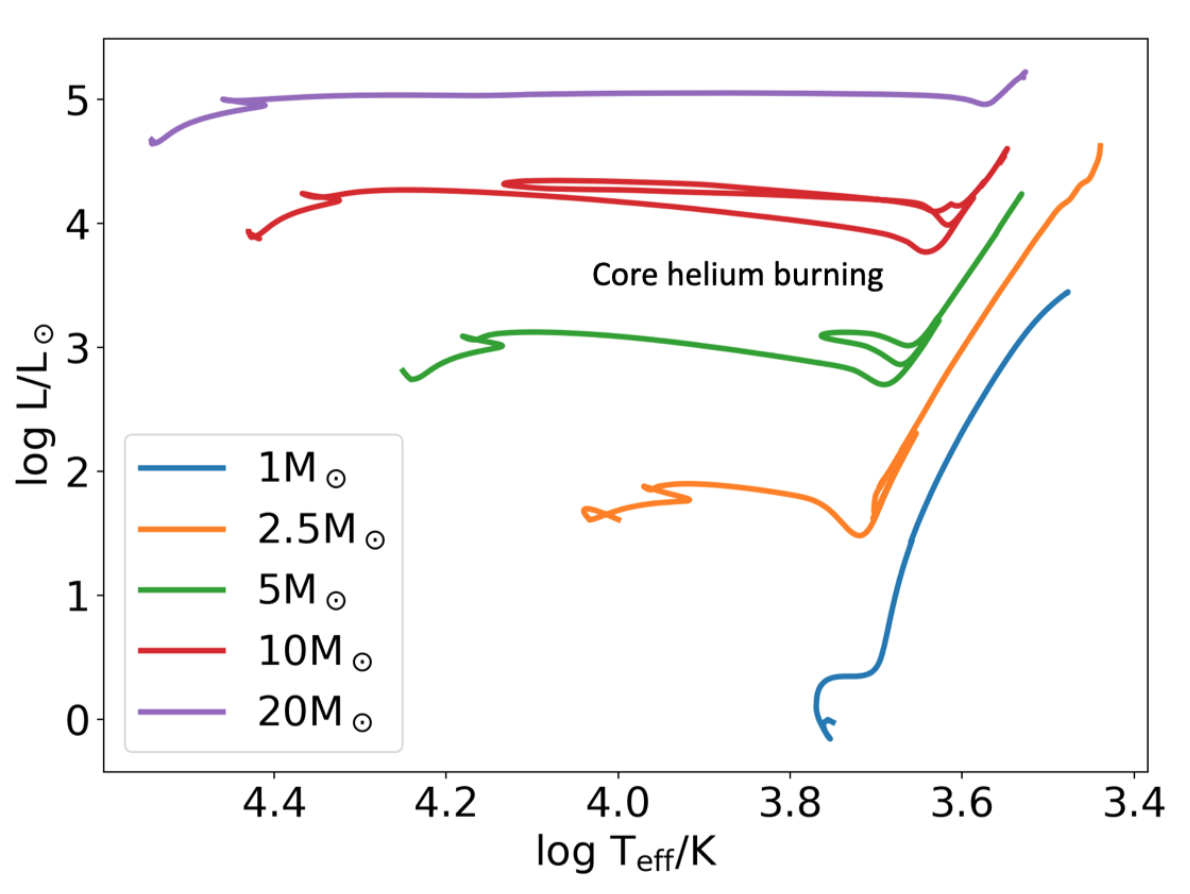}
    \caption{Example of the content covered in ThaiPASS 2019 - an HR diagram of the evolution of various masses of stars plotted with Python.}
    \label{fig:hr_diagram_2019}
\end{figure}

This portion of the summer school discussed the production of elements in stars - from low-mass stars like the Sun to massive core-collapse supernova explosions (CCSNe). Lectures covered all of the major burning regimes found in main sequence stars, and beyond to the products of stellar collapse and explosion. This material complements and extends the material covered in some A-level courses - for example the optional AQA astrophysics module which covers the Hertzsprung-Russell (HR) diagram (which also featured at ThaiPASS- see section \ref{programming_topics}) and supernova. Figure \ref{fig:hr_diagram_2019} shows HR diagram tracks for various masses of stars, an example of some of the material that was covered during the two ThaiPASS summer schools.

The students were then presented with a data-mining task, in which they were asked to investigate stellar models computed by the NuGrid collaboration \cite{pignatari2016nugrid}. They were asked to find various points in the evolution of a star, and show how the abundances of various isotopes changed as the star evolved using published tools \cite{ritter2016nupycee}. The nugridpy python package was used by the students, in the same way that it is used to interface with real data-mining research. This parallels the work of IRIS (Institute for Research in Schools) which promotes cutting edge research in UK schools in collaboration with UK universities \cite{hatfield2019iris,parker2017real}.


\subsubsection{Galactic Chemical Evolution and Galactic Dynamics}

Whilst not directly addressed in A-level courses, the natural progression after discussing how isotopes and elements are produced in stars is to ask how those products are distributed around the galaxy, and indeed how a galaxy forms. A series of lectures was delivered in the 2018 ThaiPASS school addressing this, and the effects of supernova explosions on the formations of galaxies were discussed. Students were then trained in using OMEGA \cite{cote2018omega} a one dimensional galactic chemical evolution code, stellar data library and data handling package which is used in current GCE research. Students were tasked with fitting the evolution of different elements in the Milky Way through time, by adjusting the ratio of different stars and the products of burning from these. Students enjoyed this task and showed good progress in using the code to achieve the expected evolution of a number of key elements.


\subsubsection{Galaxies and dark matter}

Two lectures were given regarding galaxy rotation curves (orbital velocity as a function of galactic radius) and how the discrepancy between theoretical predictions and observed rotation curves led to the theory of dark matter. Though these topics are not covered in A-level syllabuses, the level of the lectures was not beyond that of an A-level student. 

The task consisted of three parts, which drew on python skills learned in previous workshops during the week. Specifically, the tasks required data handling, graph plotting and defining functions. Students were provided with rotation curve data for the galaxy M31 (Andromeda) obtained from \cite{2018PASJ...70...31S} and density profiles calculated from \cite{2012A&A...546A...4T}. Task one required students to plot the rotation curve using the data from \cite{2018PASJ...70...31S}; task two was to plot the theoretically predicted rotation curve from only baryonic matter contributions. The two graphs from tasks one and two would not match, which indicates baryonic matter is not sufficient to explain observations. Task three was then intended for students to fix the theoretical rotation curve by adding a dark matter component. 

The purpose of the task was to consolidate the programming skills learned over the week by presenting a guided enquiry-based task. The physics of the task was signposted (and the data curated to avoid complexity that would distract from the overall aims) whereas the method of programming was left to the student. 

Tasks one and two, which involved loading and plotting data along with a simple calculation, were completed by all students with little to no help required. Task three proved more difficult. Most students approached the task correctly from a programming perspective; however, the increased conceptual and mathematical complexity gave rise to erroneous results. 
\subsubsection{Supernova}

This topic covered the ends of the lives of massive stars, their explosions as supernova and the evolution of supernova remnants. During the lecture, supernova were discussed in general including an explanation of the different types of supernova, the physics of the supernova event and examples of observations. A large part of the lecture focused on the details of supernova remnants (SNRs), discussing the three phases of the expansion of the shell of ejected material and showing examples of observations of SNRs at each phase. The analytic equations that describe the evolution of the first two phases (the free expansion and Sedov-Taylor phases) were introduced as a way to estimate the radius and velocity of the SNR as a function of time. Modeling supernova and their remnants was discussed, both using the analytical equations and hydrodynamical simulations. The end of the lecture included a brief explanation of how the evolution of galaxies is effected by supernova feedback.
 
The first task involved modeling the dynamical evolution of a SNR starting from the Sedov-Taylor phase. Using python, the students were required to use the given equations to calculate the initial kinetic energy and radius resulting from the end of the free expansion phase. Then using this as a starting point for the Sedov-Taylor phase, use the given equations to calculate the radius and velocity of the expanding shell of gas as a function of time and plot their results. The second task used a module called yt (\href{https://yt-project.org}{https://yt-project.org}) to analyse data which had been produced by a hydrodynamical simulation of a supernova blast wave. Using this they could visualise different properties (e.g. density, temperature) in “slices” through the simulation box and compare the expansion over time to that estimated from the analytical equations. As an extra task for those who had completed the first two, the students could download the plots produced from analysing the simulation data and combine them into a movie.

\subsection{Programming content} \label{programming_topics}

One of the goals of ThaiPASS is to train students in data handling. Ideally, students should have a basic knowledge of Python before they arrive, so that the week can be devoted to more complex tasks. Therefore, we developed a number of introductory `Starter Pack' materials which were provided to students before the school. These included a guide on installation of Python on their own machines, a Jupyter notebook and exercises covering basic Python commands, and an advanced notebook covering more difficult topics in Python (e.g. reading and writing a file). These materials are available online on the \href{https://github.com/Milne-Centre/ThaiPASS2018}{Milne Centre GitHub} page, and together constitute a first introduction to Python programming. For ThaiPASS'19, video tutorials\footnote{The videos were produced by Dr Teeraparb Chantavat (one of the Thai PIs for ThaiPASS), and made available as a MOOC through \url{https://thaimooc.org}.} were also provided as part of the Starter Pack.

The Starter Pack was found to have been a valuable tool for those students which engaged with them beforehand. These notebooks would be appropriate for any pupils of ages 11-18, and could be supplied to schools easily as a first introduction for students with no prior knowledge of Python.

The main goal of the programming section of ThaiPASS was to introduce the fundamental tools of Python, and some of the ways in which those tools could be applied in the context of data-intensive astronomy. Although it was not our primary intention to give a comprehensive course on computer programming, the skills required for the  set tasks led to students becoming much more familiar and confident in their application of programming skills.

The Python tutorials and sessions during the summer schools covered a variety of tasks - producing HR diagrams of sample spectroscopic data, calculating the mechanics of orbital motion, handling large data sets during the GCE, nucleosynthesis tasks and galactic dynamics sessions, and further skills such as producing function, using loops to complete tasks and more. These tasks are largely modular, and lend themselves to being changed between sessions to suite the interests and capabilities of the students.


\subsubsection{Data handling and plotting with Python}

\begin{figure}
    \centering
    \includegraphics[width=0.7\textwidth]{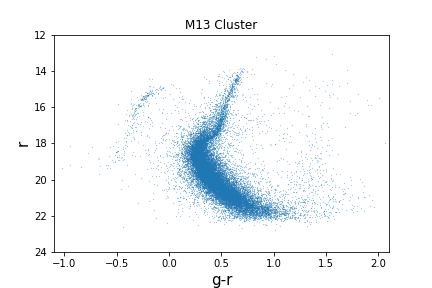}
    \caption{Sample python plot of the HR diagram of the M13 cluster. All students were able to produce a similar plot with varying degrees of assistance.}
    \label{fig:simple_python_M13}
\end{figure}

One of the more common introductory topics for astrophysics is the Hertzsprung-Russell (HR) diagram.
Colour-magnitude diagrams are directly related to these HR diagrams, which show the luminosity (or brightness) of an object as a function of its colour (a proxy for temperature). 
These are fundamental tools in observational astronomy and allow classification of stars by their evolutionary stages. 
Therefore, we provided the students with a sample of real observational data obtained from the Sloan Digital Sky Survey (SDSS)\cite{York2000} SEGUE value added catalogue, which contains a mixture of well-studied open and globular stellar clusters\cite{An2008}.
Students were tasked with replicating the colour-magnitude diagrams for each stellar cluster from this sample in order to illustrate the observed differences between open and globular cluster systems.
Figure \ref{fig:simple_python_M13} shows an example Python plot of the M13 cluster using the SDSS data supplied to the students. 

This task calls on a wide range of data-handling skills such as accessing data in a file, manipulating that data, and finally presenting it in a legible way, all of which are key skills for any fields involving data. Students took to this task well, with all participants being able to produce their own colour magnitude plot from the data provided, though varying degrees of help were required. By comparison with any of the widely available images of M13 and its accompanying HR diagram, Python plotting tools produce publication-quality figures. Students were able to produce plots similar to that shwon in Figure \ref{fig:simple_python_M13} after only a few hours tuition on the first day, and were instructed to verify their work by comparing with any published HR diagrams of M13 that they could find.

\subsubsection{Forward Euler Method}

Another computational topic covered at ThaiPASS was producing a Forward-Euler solver - a numerical method for solving ordinary differential equations (for a full discussion see, for example, \cite{atkinson2008introduction, ascher1998computer}). Because of its complexity, this task lends itself well to discussing the idea of building an algorithm, and the idea of planning out what needs to be coded up. Future ThaiPASS events may extend this task further to allow students the opportunity to experience algorithm development and gain proficiency with planning code methodology as it is well suited to this application.

Numerical modelling of ordinary differential equations are the backbone of many codes used in computational astrophysics and a key tool in all of the computational work undertaken in STEM subjects. In both ThaiPASS'18 and 19, students were tasked with producing their own numerical solver for integrating ordinary differential equations.

This task is particularly difficult, and was set with the expectation that students would work collaboratively. Most students succeeded, with assistance either from peers or the teaching staff, in the programming of the forward Euler solver, a particularly impressive feat considering the difficulty. The full task of modelling the orbits of an example planet was completed only by a few students, however the difficulty of this task is more appropriate for a university level course with undergraduates at the University of Hull performing similar computational tasks in years 2 and 3. This task was largely included as a stretch and challenge goal for the most capable. Based on the experience built in ThaiPASS 2018, the 2019 ThaiPASS summer school, included a brief introduction to basic calculus (as finding the area under a curve) which was not known by all students at the start of the summer school. We noted significant increase in attainment and engagement from students with the addition of this material. Future introductory material should include either a basic description of the principals of calculus or a thorough description of the Euler method if this is to be retained. Another option to increase confidence in participating students would be to make available a model answer with solutions., which they may then compare to in order to verify their results.

\subsection{STEM-career content}

Each ThaiPASS also featured a \textit{`University and Career Day'}, with the aim of inspiring students to take up STEM education at university level and perhaps as a career. 

During the event, Thai professionals (STEM graduates)  gave inspirational and interactive talks on ways in which STEM subjects can be taken further at university level and as a career, and the career paths available to graduates with STEM skills. Previous lectures include: `\textit{How to become an astronomer?}' and  `\textit{What do astronomers do?}' (both by Thai astronomy researchers) as well as `\textit{Why STEM? Career paths of STEM graduates}' by a speaker from the British Council. 

\section{Lessons learned from ThaiPASS}
\label{analysis}

We now discuss the impact of the program, as based on surveys of the participants and discuss some of the challenges faced in presenting the material.

\subsection{Impact of ThaiPASS} \label{student_teacher_feedback}

At the end of each five-day summer school, participants were asked to complete an anonymous questionnaire in order to assess the impact of the summer school and to improve the next iteration of ThaiPASS. We asked the students to rate their experiences of the various activities, and the usefulness of those activities.

\subsubsection{Results from 2018}

Results for each of the questions are presented below:

\begin{figure}
     \centering
     \begin{subfigure}[b]{0.45\textwidth}
         \centering
         \includegraphics[width=\textwidth]{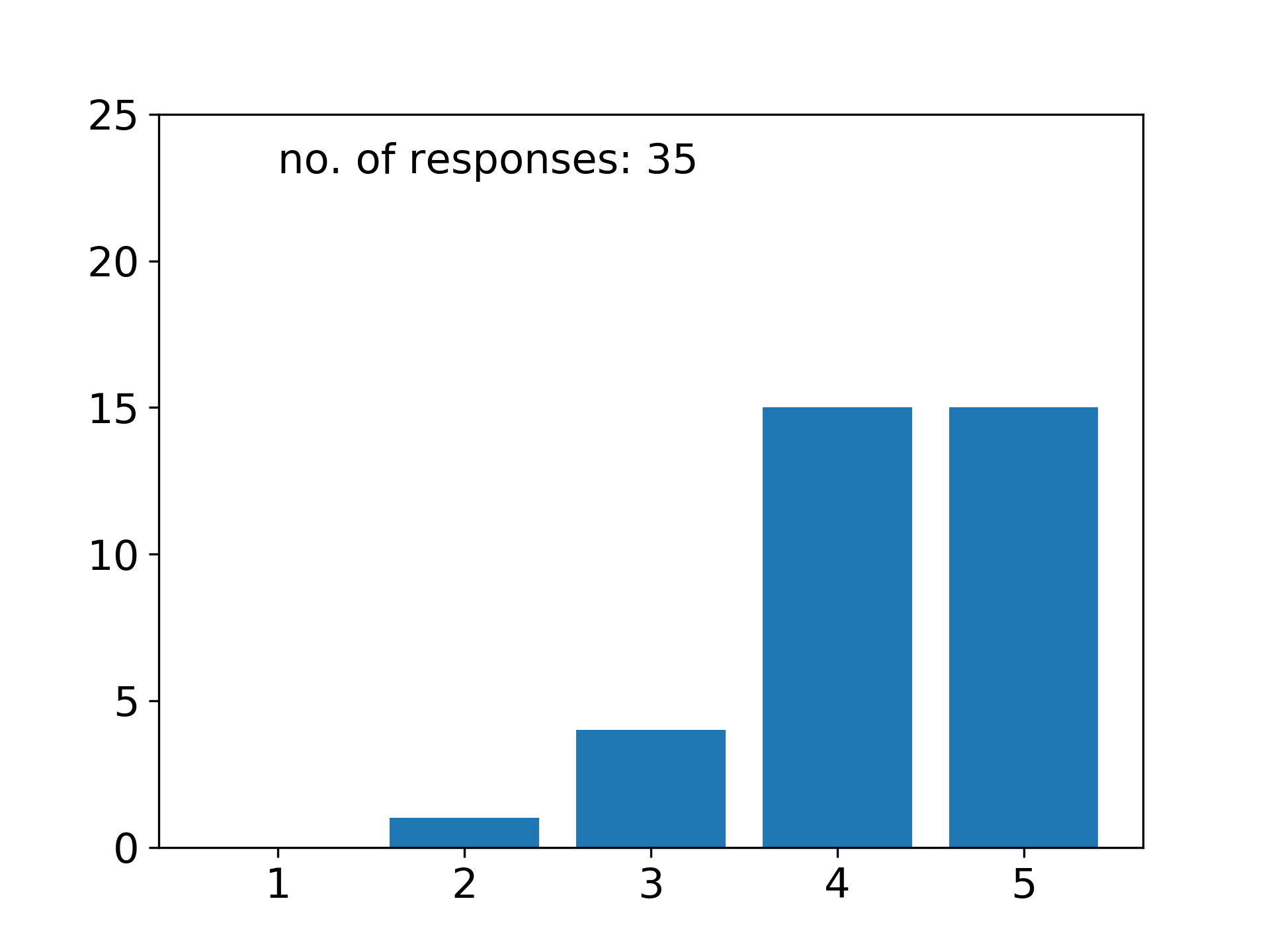}
         \caption{Did you enjoy the lecture and activities related to Kepler's laws of motion and the Hertzsprung-Russel Diagram?}
         \label{fig:Q2a_plot}
     \end{subfigure}
     \hfill
     \begin{subfigure}[b]{0.45\textwidth}
         \centering
         \includegraphics[width=\textwidth]{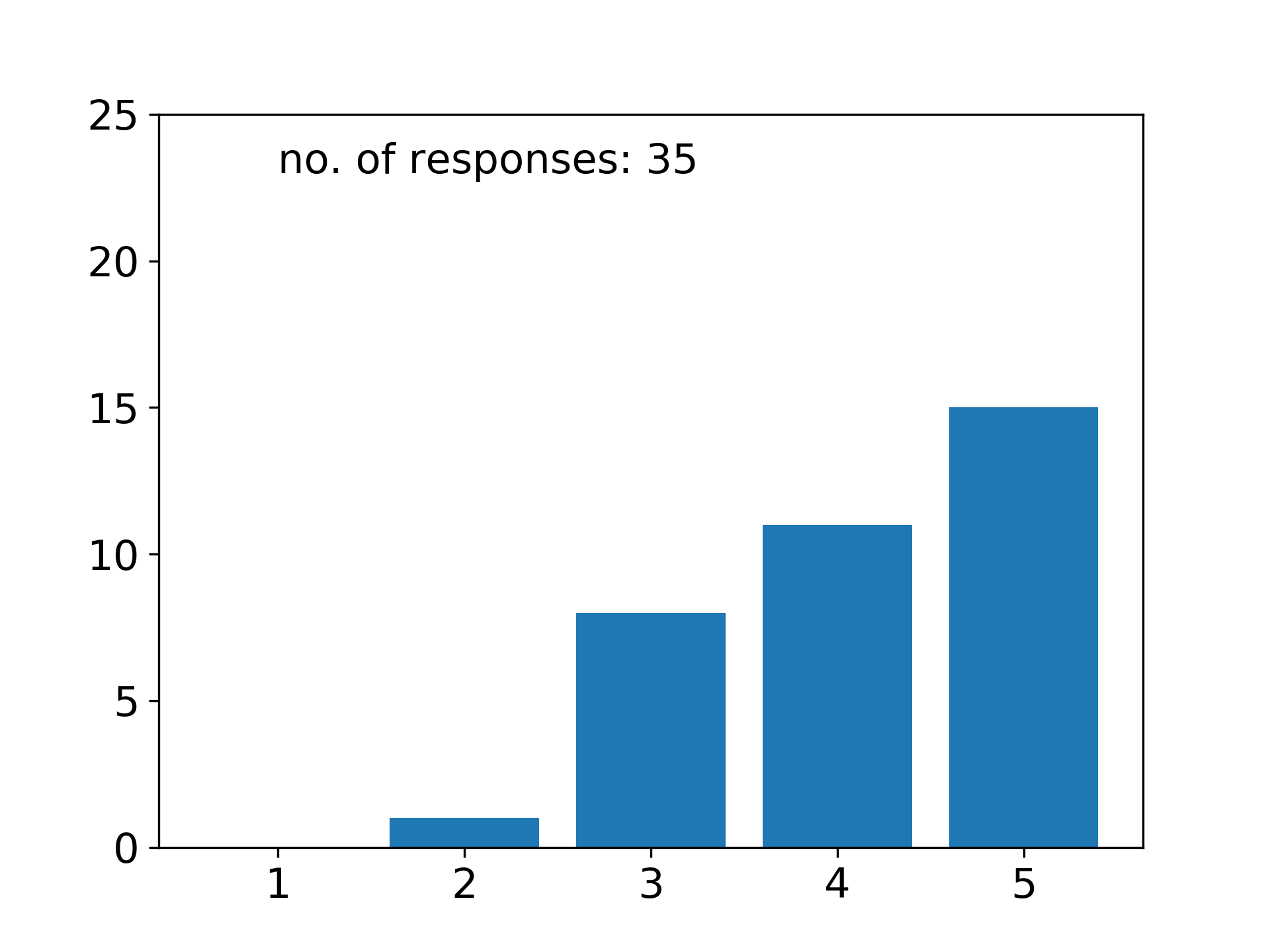}
         \caption{Did you enjoy the session focusing on nuclear astrophysics and the production of the elements?}
         \label{fig:Q2b_plot}
     \end{subfigure}
     \hfill
     \begin{subfigure}[b]{0.45\textwidth}
         \centering
         \includegraphics[width=\textwidth]{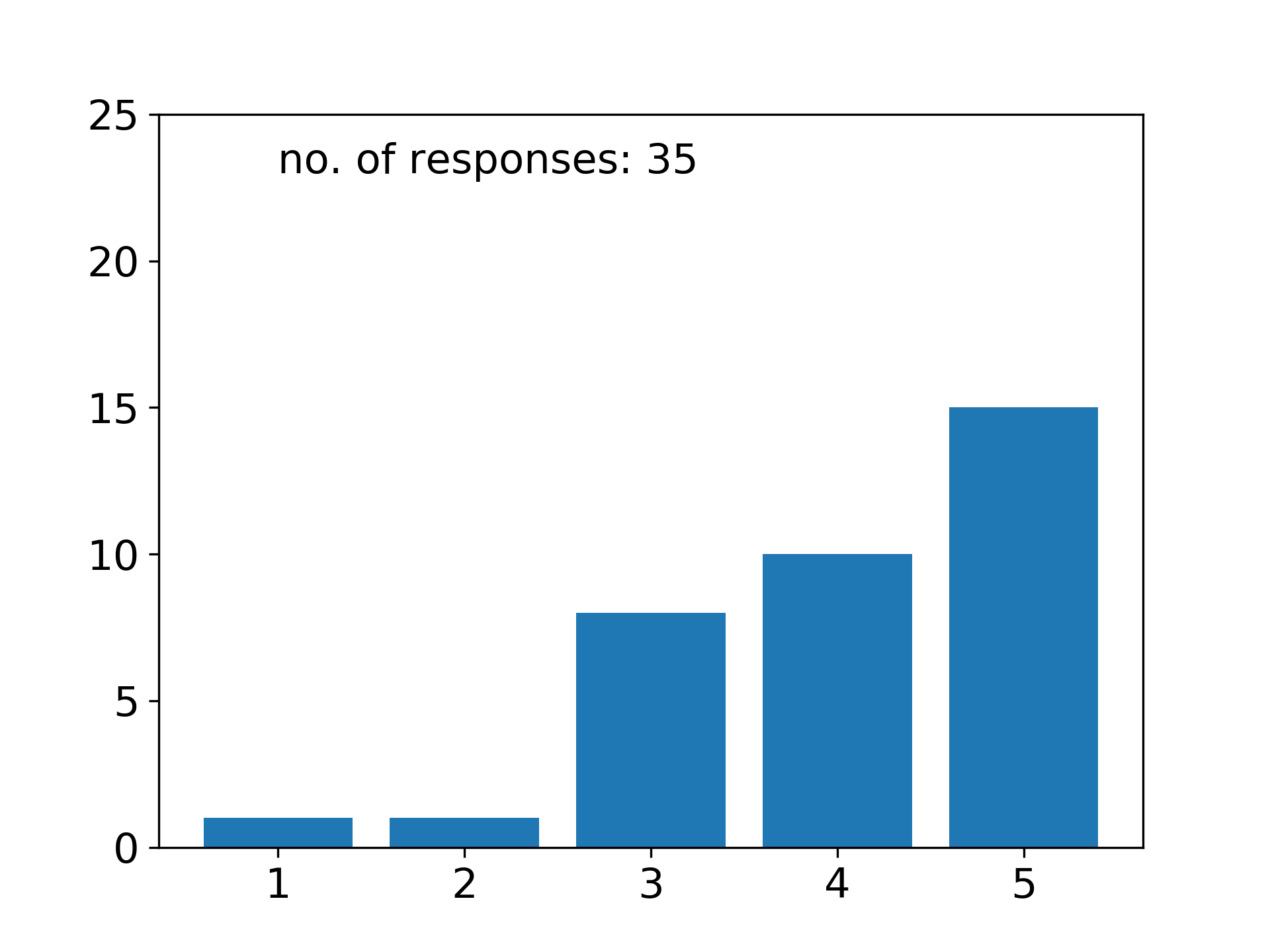}
         \caption{Did you enjoy the session focusing on the Euler method and solving ODE?}
         \label{fig:Q2c_plot}
     \end{subfigure}
     \hfill
     \begin{subfigure}[b]{0.45\textwidth}
         \centering
         \includegraphics[width=\textwidth]{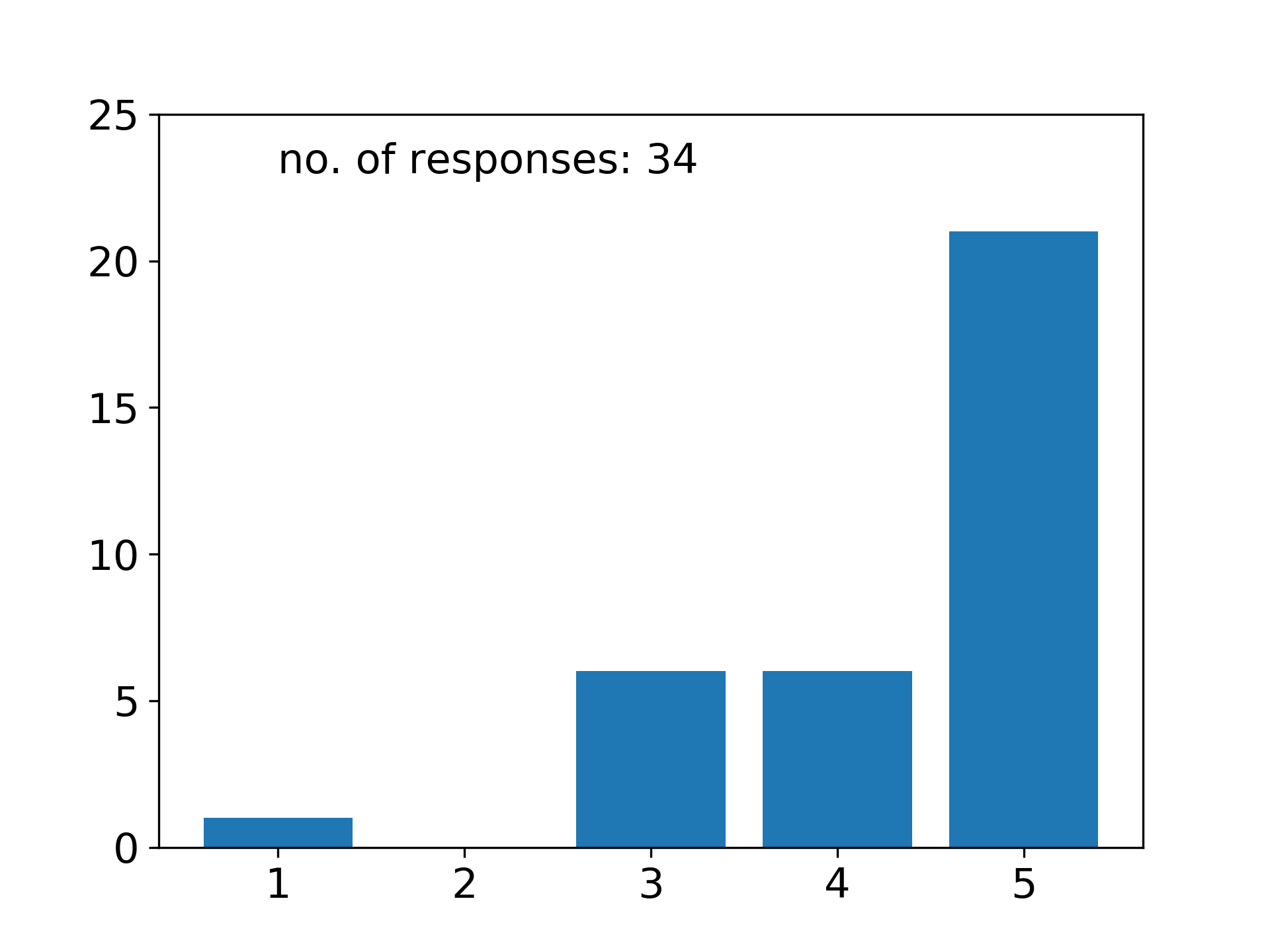}
         \caption{Did you enjoy the galactic chemical evolution lectures and activities?}
         \label{fig:Q2d_plot}
     \end{subfigure}
        \caption{Student responses of how much they enjoyed each day's subject material, with 1 being `not at all' and 5 being `very much'. }
        \label{fig:all_Q2_results}
\end{figure}

\begin{figure}
     \centering
     \begin{subfigure}[b]{0.45\textwidth}
         \centering
         \includegraphics[width=\textwidth]{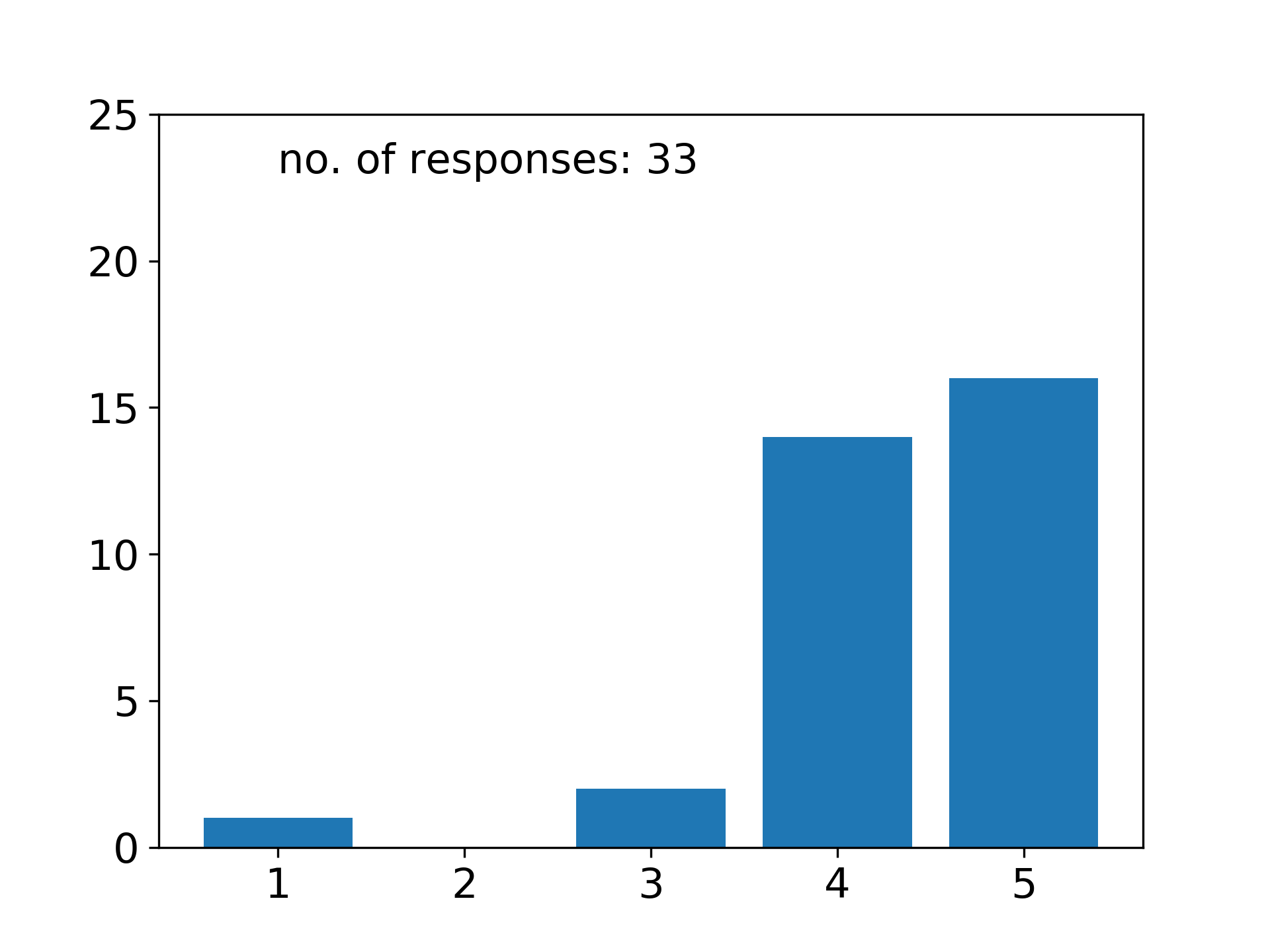}
         \caption{Has the summer school increased your knowledge of Python syntax and application?}
         \label{fig:Q4a_plot}
     \end{subfigure}
     \hfill
     \begin{subfigure}[b]{0.45\textwidth}
         \centering
         \includegraphics[width=\textwidth]{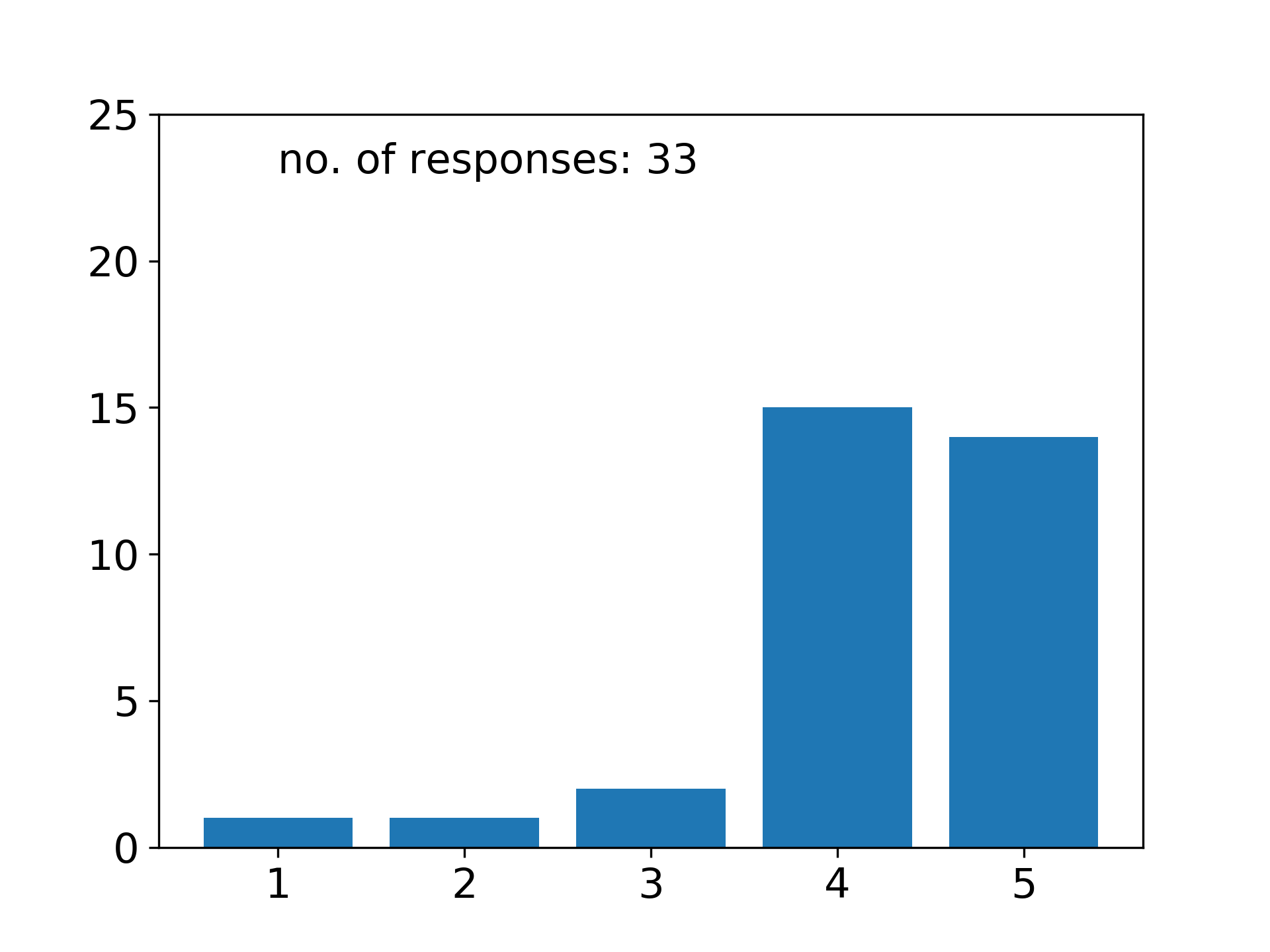}
         \caption{Has the summer school increased your understanding of the application of computer science?}
         \label{fig:Q4b_plot}
     \end{subfigure}
     \hfill
     \begin{subfigure}[b]{0.45\textwidth}
         \centering
         \includegraphics[width=\textwidth]{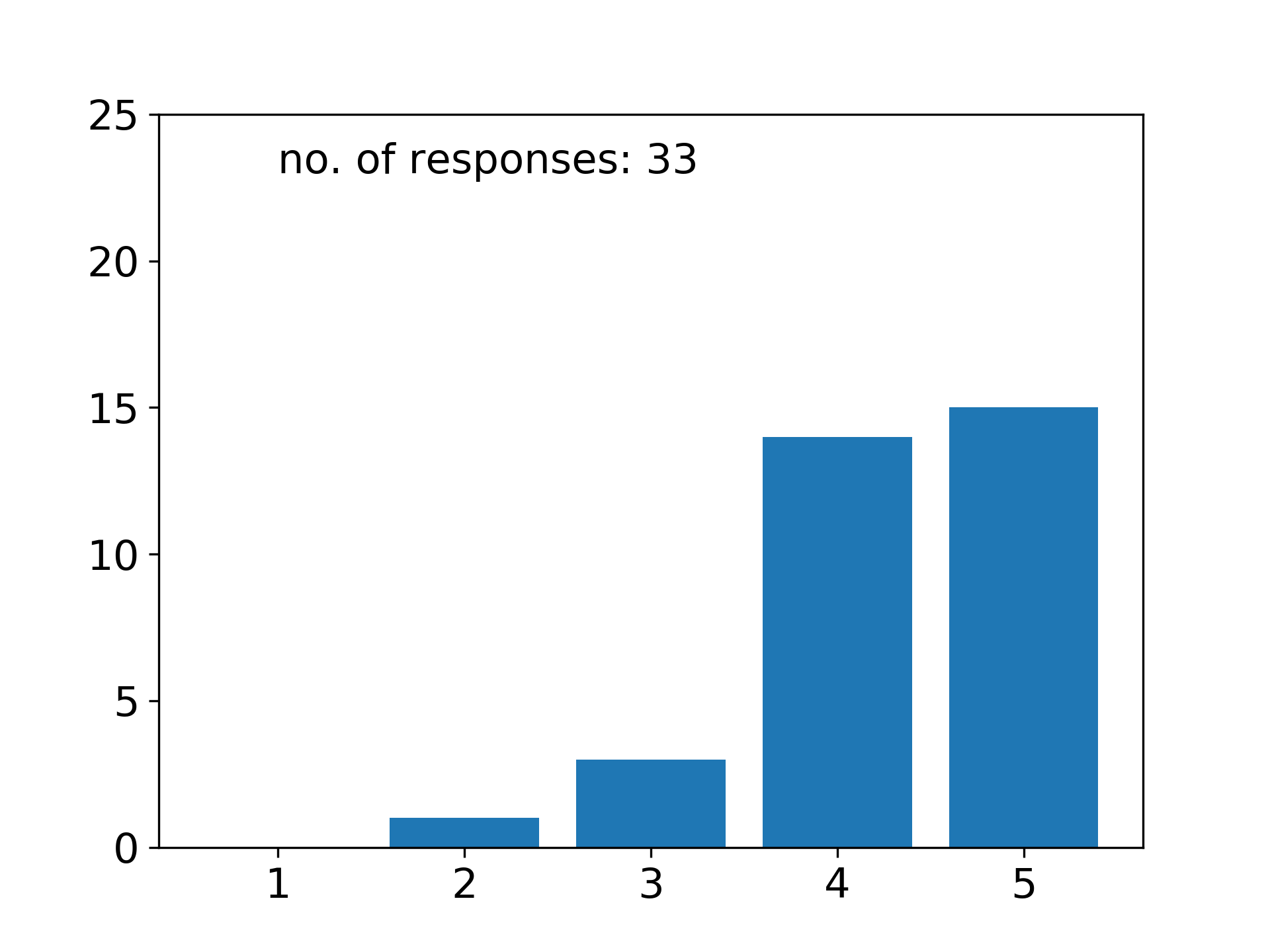}
         \caption{Enjoyment of lectures}
         \label{fig:Q4c_plot}
     \end{subfigure}
     \hfill
     \begin{subfigure}[b]{0.45\textwidth}
         \centering
         \includegraphics[width=\textwidth]{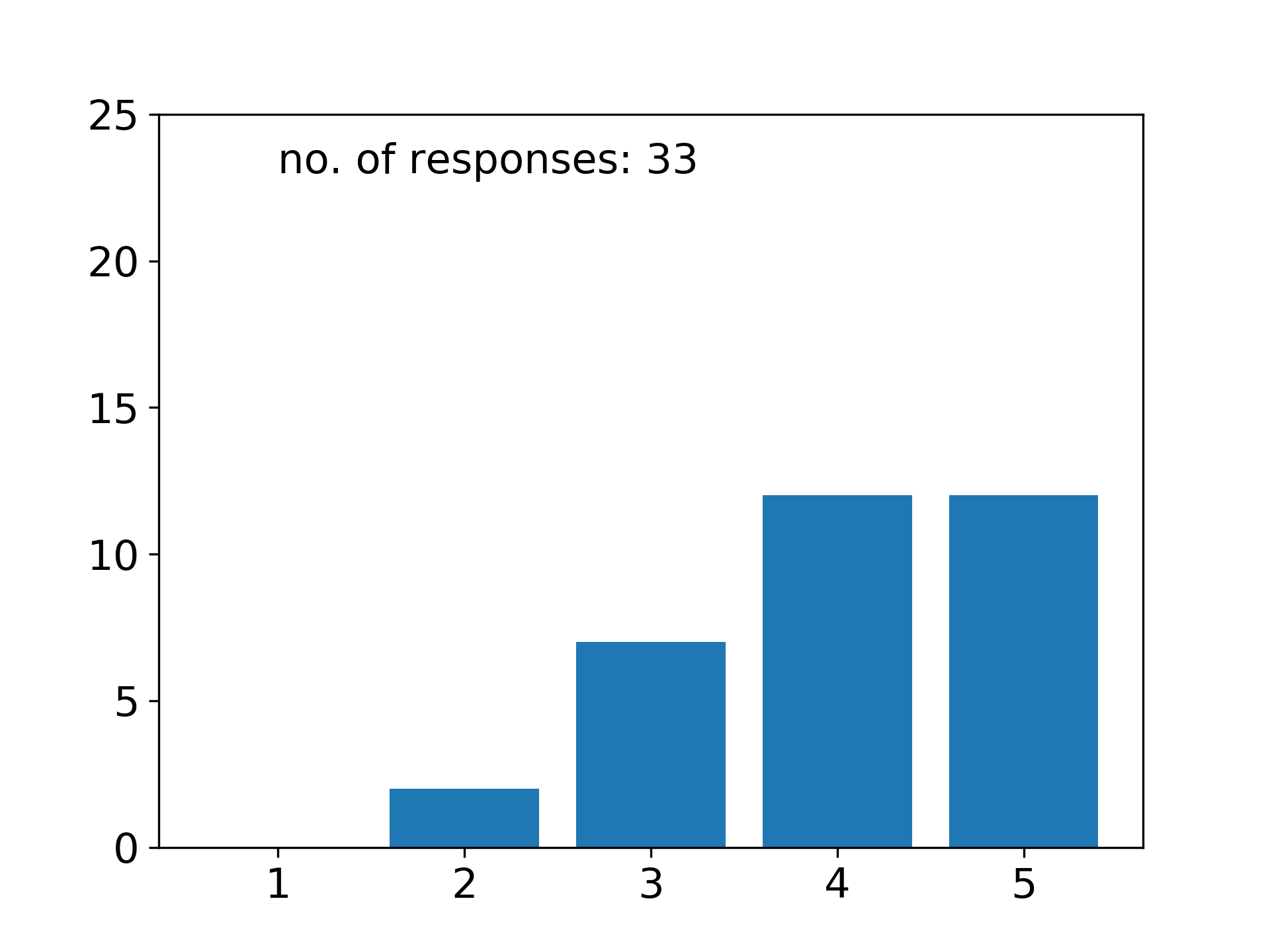}
         \caption{Did you enjoy the coding sessions?}
         \label{fig:Q4d_plot}
     \end{subfigure}
     \hfill
     \begin{subfigure}[b]{0.45\textwidth}
         \centering
         \includegraphics[width=\textwidth]{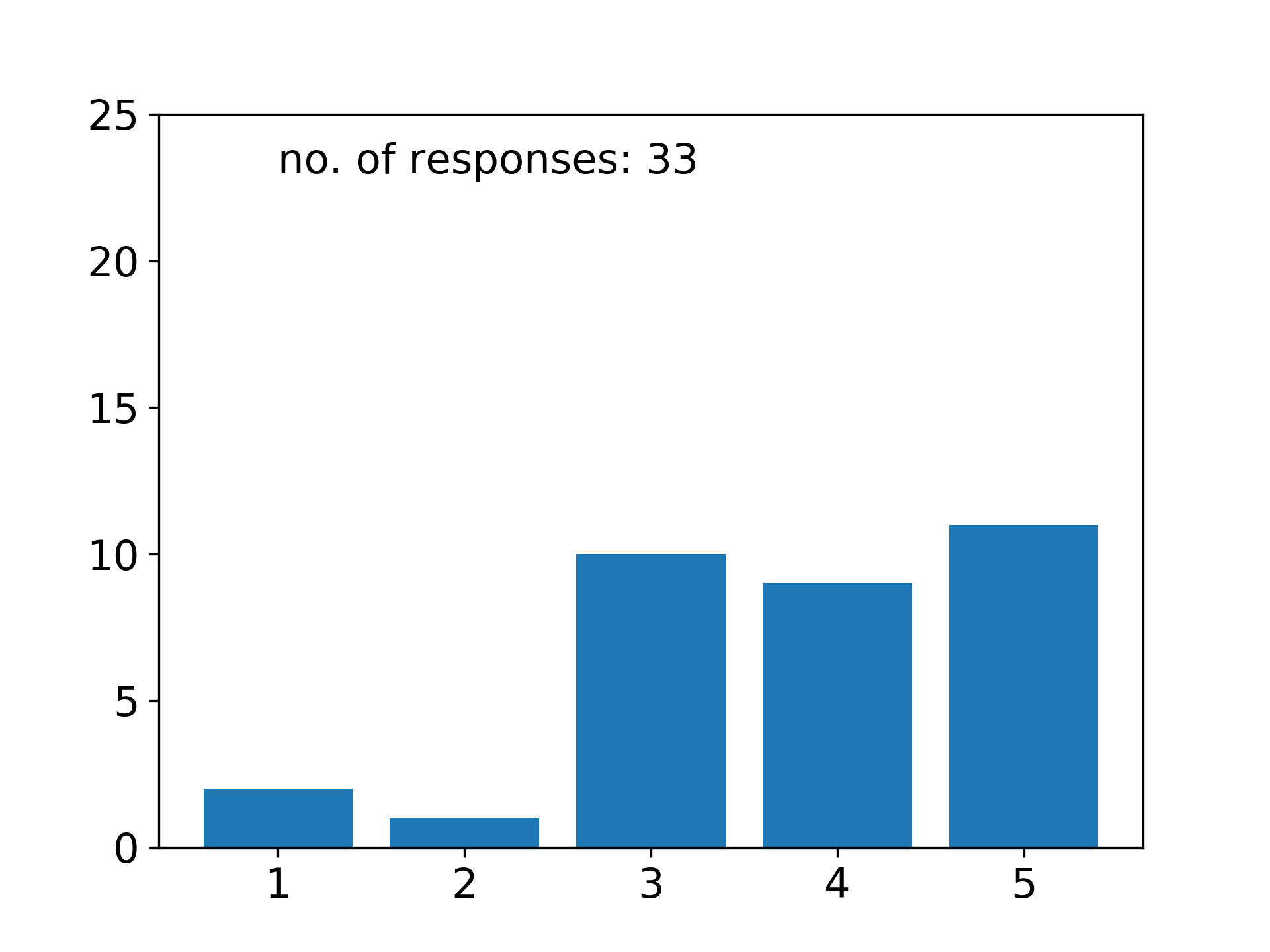}
         \caption{How confident are you plotting with python after the summer school?}
         \label{fig:Q4e_plot}
     \end{subfigure}
     \hfill
     \begin{subfigure}[b]{0.45\textwidth}
         \centering
         \includegraphics[width=\textwidth]{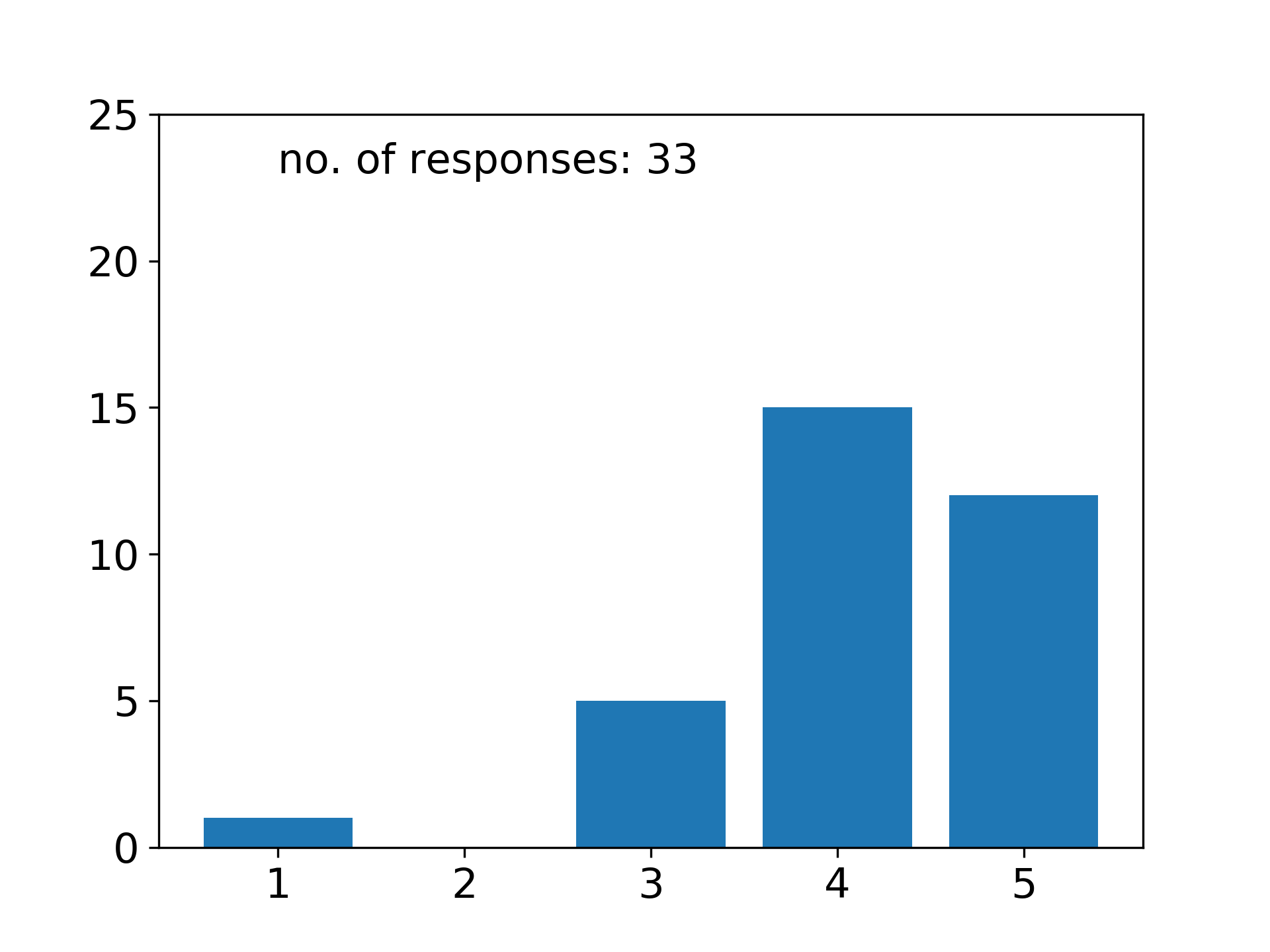}
         \caption{Would you now study a STEM subject at university?}
         \label{fig:Q4f_plot}
     \end{subfigure}
        \caption{Student response to questions, on a scale of 1 (worst) to 5 (best).}
        \label{fig:all_Q4_results}
\end{figure}

As can be seen from Figures \ref{fig:all_Q2_results} and \ref{fig:all_Q4_results}, the feedback from the week long school is largely positive. Figure \ref{fig:all_Q2_results} shows student enjoyment of the various activities presented during the 2018 summer school on a scale of one to 5, 5 being the most positive response. Overwhelmingly students enjoyed the activities. The Euler method received the worst response however the majority of students still rated this activity from average to good. The galactic chemical evolution lectures are clearly the most well received. The material for this session was accessible, students were engaged, and progress was identifiable in the tasks that they were set. Future ThaiPASS summer school lecture material should also be accessible, engaging and relevant, with measurable outcomes for student progression which they can appreciate at the end of the course. Further development of teaching materials is underway to facilitate this.

Figure \ref{fig:all_Q4_results} asks students to rate their experience according to the various questions outlined. As is shown in (a) and (b), almost all students report a large increase in their technical skills in Python and the applications of computer science more generally. Almost all students also rate the enjoyment of lectures highly too, showing that material is engaging to the target audience. The response to the increase in  programming and plotting skills are more varied, however this is likely due to the difficult nature of the material, and the response still shows that students believe their skills have improved significantly. The likelihood of students studying a STEM subject further is also encouragingly high.


\subsubsection{Results from 2019}

For ThaiPASS'19, some changes to the questionnaire were made, and a pre- and post-school survey were undertaken with the intention of trying to better measure the impact of the school. The full questionnaire is provided in the Appendix.

We see from Figure \ref{fig:responses2019} that students have engaged with the lecture material and found it useful and informative, feedback from NARIT (National Astronomical Research Institute of Thailand) and teaching staff is also very positive. Students were generally well disposed to the subject material before the school, with most giving a score of 4 or 5. The post-school surveys showed a higher proportion of 5s, suggesting we have helped to enhance students' perceptions. Particularly gratifying is the response to the question `Are you interested in a STEM career?': after the school, the student response was overwhelmingly of the highest response. While students said they enjoyed both the astronomy lectures and the coding sessions, they expressed a clear preference for the hands-on activities. This suggests it may be beneficial to either change the balance of lectures to activities, or to intersperse the two so as not to lose students' interest.

\begin{figure}
    \centering
    \includegraphics[width=\columnwidth]{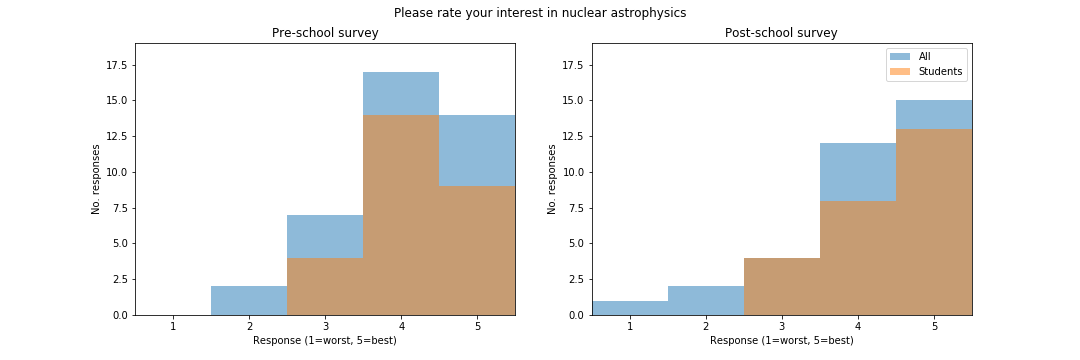}
    \includegraphics[width=\columnwidth]{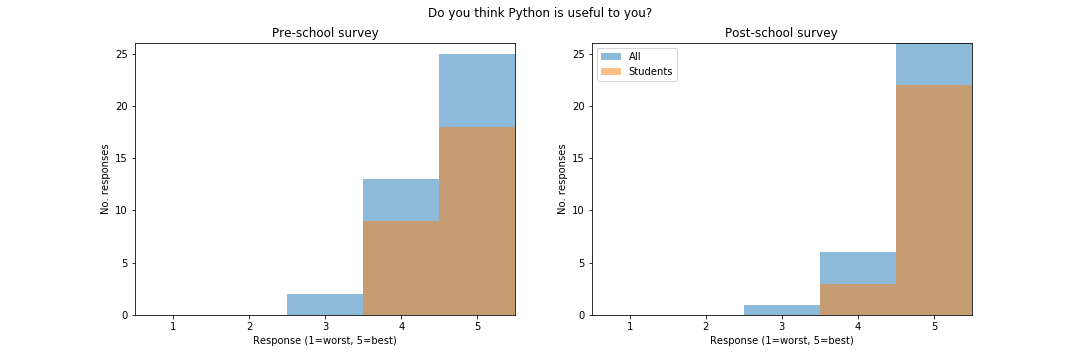}
    \includegraphics[width=\columnwidth]{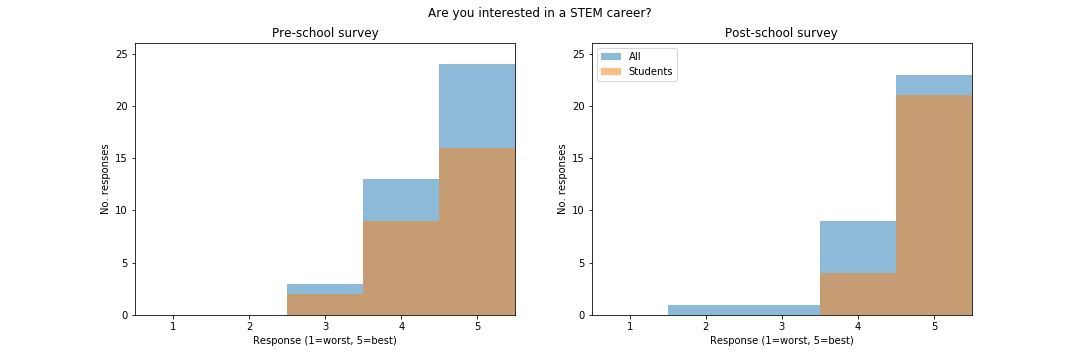}
    \caption{Some of the questions and responses from the pre- and post-school surveys conducted during ThaiPASS 2019. All responses includes questionnaires from teachers and NARIT participants.}
    \label{fig:responses2019}
\end{figure}

\subsection{Challenges}
\label{challenges}

Although ThaiPASS'18 and 19 were delivered without major difficulties, it is worth discussing two aspects which are important to consider for those organising outreach events, especially international ones.

\subsubsection{Language barrier} 

Although English is a compulsory subject in all Thai schools, participants came to us with various levels of skill in English. Most students had little trouble comprehending written instructions in English, although oral communication was occasionally challenging. It was helpful to have Thai instructors on hand in the practical sessions to help lessen the language barrier. Perhaps the most important mitigating measure was The Starter Pack which ensured that participants were familiar with key Python and astronomical terms in English before arrival. A good number of students were able to enthusiastically communicate with the UK team and asked interesting questions (both orally and through  \href{http://www.mentimeter.com}{Mentimeter} an anonymous online forum).

\subsubsection{Gender balance}

ThaiPASS participation is based on nominations from schools by teachers. From the very beginning, we have always strongly encouraged each school to nominate at least one female student to attend ThaiPASS (unless the school is single gender). Tables \ref{student_gender}-\ref{instructor_gender} break down the participating  students, teachers and instructors respectively by gender, for the 2018 and 2019 workshops.

\begin{table}
    \centering
    \begin{tabular}{|c|c|c|c|}
    \hline
         & Male & Female  & M:F ratio \\
         \hline
       ThaiPASS'18  & 15 & 15 & 50:50 \\
       ThaiPASS'19	& 20 & 10 & 67:33 \\
       \hline
    \end{tabular}
    \caption{Student participants by gender.}
    \label{student_gender}
\end{table}

\begin{table}
    \centering
    \begin{tabular}{|c|c|c|c|}
    \hline
         & Male & Female  & M:F ratio \\
         \hline
       ThaiPASS'18  & 6 & 4 & 60:40 \\
       ThaiPASS'19	& 7 & 2 & 78:22 \\
       \hline
    \end{tabular}
    \caption{Teacher participants by gender.}
    \label{teacher_gender}
\end{table}

\begin{table}
    \centering
    \begin{tabular}{|c|c|c|c|}
    \hline
         & Male & Female  & M:F ratio \\
         \hline
       ThaiPASS'18  & 6 & 2 & 75:25 \\
       ThaiPASS'19	& 6 & 4 & 60:40 \\
       \hline
    \end{tabular}
    \caption{Instructors by gender.}
    \label{instructor_gender}
\end{table}

\noindent
Although ThaiPASS'19 performed worse in terms of overall F:M ratio, we were able to award `outstanding participant' prizes to 3 male and 3 female students. In terms of speakers, we have had a slight increase over the past two years, though more work needs to be done to achieve an even balance. Feedback from female participants clearly praised the positive impact of having female instructors, who were regarded as positive role models for women in STEM. A female participant at ThaiPASS'19 expressed this as follows:

\begin{quote}
    ``\textit{I got a chance to see my dream self. I have a dream to be a woman who graduated with a doctoral degree. After seeing Dr. Claire’s lectures, I have a feeling that this is possible. She looks special and even more extraordinary when she is talking about something educational; which inspired me to try harder to become like that.}"
\end{quote}

\noindent
Funding received for ThaiPASS'20 and '21 allows us to plan for a far bigger event. We therefore hope for a much larger pool of applicants from which we hope to be able to draw a better gender balance. We will explicitly ask schools to nominate female students, as we have done with the past schools. We will also try to recruit more female instructors, particularly when recruiting Ph.D. assistants, with the aim of achieving a 50:50 balance. 

Finally, because teachers were not the primary target audience for the 2018-19 workshops, we did not monitor the gender balance nor actively encouraged participation from female teachers. However, the next ThaiPASS will feature a separate 5-day Python training for teachers (around 20 expected), and we will work to improve the gender balance within this group.

\section{Plans for impact in the UK}
\label{plans}


A high-impact legacy of the ThaiPASS project is for a similar summer school to be held in the UK, particularly in Hull. Hull is an area with 47 Lower layer Super Output Areas (LSOA) in the bottom 20\% nationally \cite{prothero_2016}, the highest national proportion of deprived areas \cite{ministry_of_housing_2019}, and remains one of the cities in the UK with a highest number of under-performing schools. The teaching material developed for ThaiPASS can be adapted to organise UK-based events targeting under-performing schools in Yorkshire and hosted by the University of Hull. The University of Hull has a very engaged outreach team, with participation from undergraduates, postgraduates, teaching and research staff. We anticipate that the university could easily provide one summer school initiative based in Hull each year, dependent on funding. With a similar model to the ThaiPASS initiative, this could reach students in up to 30 local authority schools each year, with a cohort of 40-60 students. 

ThaiPASS also offers an opportunity for continuous professional development for teaching professionals, who may wish to incorporate some of the material in their classrooms, if similar interest is shown in these projects then a further 20 teachers may attend each summer school. Further programs using these materials may be rolled out across the country, with one such event being held in Glasgow as part of a teacher career professional development (CPD) event. Further summer schools, teacher training events and outreach activities in targeted schools are possible with the continued engagement of academics and research students at participating universities. 

Due to the flexible nature of the materials and activities, this model may be adapted by any research team who wishes to present their data and tools to interested pupils and teachers. Our hope is that we have provided a viable framework that allows students to engage with cutting-edge research from many different universities, whilst building on topics in the GCSE and A-level syllabi. Specifically in the area of astronomy and astrophysics,  one could develop projects or workshops based on these topics, amongst others:
\begin{itemize}
    \item Telescope optics and design
    \item Spectroscopy and stellar classification
    \item Compact-object physics
    \item Cosmology
    \item Exoplanet detection.
\end{itemize}


The Physics group at Hull (particularly the Milne Centre) has a strong track-record of commitment to outreach within the Yorkshire region and beyond, and new participants from other institutes offers the possibility of producing a comprehensive suite of high-quality astronomy education materials, suitable for a broad range of students at all levels of education, and accessible to all schools no matter their performance or available funding.

Assessing the impact of these summer schools will be essential to the improvement and development of the project. Future events hosted at universities would be assessed in a similar way as the current ThaiPASS summer schools - a mixture of continuous assessment during the course of the summer school, questionnaires and written statements from participating students and teachers. If the project is developed further to be used more independently in schools and over a longer timescale, a number of other options for assessment are possible. One such method could be the instigation of a competition where students complete a project of 4-6 weeks using current research data, producing a finished suite of code used to analyse data or to model simple astrophysical systems. A similar initiative could include a symposium or poster presentation, which introduces  students to another key aspect of scientific research - the presentation of data and results. Significant progress on projects may also be evidenced through publication, similar to some of the projects run by IRIS.

Given the challenging nature of the material, currently the resources are best suited to students in KS4 and 5. However, it is our intention to produce material suitable for both younger and more mature students, with the flexibility of the material lending itself to even post A-level study.

\section{Conclusions} \label{conclusions}

We have presented here an assessment of the first two years (2018,2019) of operation of Project ThaiPASS, a Thai-UK collaborative science outreach project. From both written and oral feedback (from students and teachers), ThaiPASS was an extremely positive and enriching experience for the participants, in addition to being a unique learning opportunity for us as education providers. The teaching resources developed for ThaiPASS are now publicly available and can be used as an introduction to one of the fastest growing programming languages in the world. These materials are also an excellent primer to a number of inter-disciplinary topics, including nuclear astrophysics and the chemical evolution of galaxies. Astronomy topics covered are also relevant to and expand upon the syllabi of various UK exam boards. This suite of resources provides both an introduction to new techniques and skills in programming, as well as extension material which can stretch and challenge even the most gifted students, whilst remaining relevant to their studies.

\section*{Acknowledgments}

The authors gratefully acknowledge the financial support of the Science \& Technology Facilities Council (STFC), through the awards of the Project ThaiPASS (ST/R006547/1) and 
of the University of Hull Consolidated Grant (ST/R000840/1). We thank the Thai PIs (Dr. Teeraparb Chantavat and Dr. Apimook Watcharangkool) for their collaboration on ThaiPASS. We also thank Dr. Kevin
Pimbblet for helpful advice on the gathering of appropriate and relevant feedback.
For the scientific data and access provided we acknowledge the support of the NuGrid collaboration, the National Science Foundation (NSF, USA) under grant No. PHY-1430152 (JINA Center for the Evolution of the Elements), from the "Lendulet-2014" Program of the Hungarian Academy of Sciences (Hungary), from the ERC Consolidator Grant (Hungary) funding scheme (Project RADIOSTAR, G.A. n. 724560) and the University of Hull High Performance Computing Facility {\sc viper}. We thank  Prof. Falk Herwig and the University of Victoria for their support and the access provided to their outreach platforms within the Cyberhubs initiative. 

\section*{Ethical Statement}

This investigation carried out in accordance with the principles outlined in the IoP ethical policy. Students' responses were collected anonymously, and consent to participate in the summer school was required from parents or legal guardians for all students.

\section*{References}
\bibliographystyle{unsrt}
\bibliography{ThaiPASS}

\begin{thebibliography}{10}

\bibitem{raddick2009galaxy}
M~Jordan Raddick, Georgia Bracey, Pamela~L Gay, Chris~J Lintott, Phil Murray,
  Kevin Schawinski, Alexander~S Szalay, and Jan Vandenberg.
\newblock Galaxy zoo: Exploring the motivations of citizen science volunteers.
\newblock {\em arXiv preprint arXiv:0909.2925}, 2009.

\bibitem{herwig2018cyberhubs}
Falk Herwig, Robert Andrassy, Nic Annau, Ondrea Clarkson, Benoit C{\^o}t{\'e},
  Aaron D’Sa, Sam Jones, Belaid Moa, Jericho O’Connell, David Porter,
  et~al.
\newblock Cyberhubs: Virtual research environments for astronomy.
\newblock {\em The Astrophysical Journal Supplement Series}, 236(1):2, 2018.

\bibitem{pignatari2016nugrid}
M~Pignatari, F~Herwig, R~Hirschi, M~Bennett, G~Rockefeller, C~Fryer, FX~Timmes,
  C~Ritter, A~Heger, S~Jones, et~al.
\newblock Nugrid stellar data set. i. stellar yields from h to bi for stars
  with metallicities z= 0.02 and z= 0.01.
\newblock {\em The Astrophysical Journal Supplement Series}, 225(2):24, 2016.

\bibitem{ritter2018nugrid}
C~Ritter, F~Herwig, S~Jones, M~Pignatari, C~Fryer, and R~Hirschi.
\newblock Nugrid stellar data set--ii. stellar yields from h to bi for stellar
  models with m zams= 1--25 m⊙ and z= 0.0001--0.02.
\newblock {\em Monthly Notices of the Royal Astronomical Society},
  480(1):538--571, 2018.

\bibitem{ritter2016nupycee}
Christian Ritter and Benoit C{\^o}t{\'e}.
\newblock Nupycee: Nugrid python chemical evolution environment.
\newblock {\em ascl}, pages ascl--1610, 2016.

\bibitem{cote2018origin}
Benoit C{\^o}t{\'e}, Chris~L Fryer, Krzysztof Belczynski, Oleg Korobkin,
  Martyna Chru{\'s}li{\'n}ska, Nicole Vassh, Matthew~R Mumpower, Jonas
  Lippuner, Trevor~M Sprouse, Rebecca Surman, et~al.
\newblock The origin of r-process elements in the milky way.
\newblock {\em The Astrophysical Journal}, 855(2):99, 2018.

\bibitem{hatfield2019iris}
Peter Hatfield, W~Furnell, A~Shenoy, E~Fox, B~Parker, L~Thomas, and EAC
  Rushton.
\newblock Iris opens pupils' eyes to real space research.
\newblock {\em Astronomy \& Geophysics}, 60(1):1--22, 2019.

\bibitem{parker2017real}
Becky Parker.
\newblock Real science, real classrooms.
\newblock {\em School Science Review}, 98(365):116--117, 2017.

\bibitem{cote2018omega}
Benoit {C{\^o}t{\'e}}, Brian~W. {O'Shea}, Christian {Ritter}, Falk {Herwig},
  and Kim~A. {Venn}.
\newblock The impact of modeling assumptions in galactic chemical evolution
  models.
\newblock {\em The Astrophysical Journal}, 835(2):128, 2017.

\bibitem{2018PASJ...70...31S}
Yoshiaki {Sofue}.
\newblock {Radial distributions of surface mass density and mass-to-luminosity
  ratio in spiral galaxies}.
\newblock {\em Publications of the Astronomical Society of Japan}, 70(2):31,
  March 2018.

\bibitem{2012A&A...546A...4T}
A.~{Tamm}, E.~{Tempel}, P.~{Tenjes}, O.~{Tihhonova}, and T.~{Tuvikene}.
\newblock {Stellar mass map and dark matter distribution in M 31}.
\newblock {\em Astronomy \& Astrophysics}, 546:A4, October 2012.

\bibitem{York2000}
Donald~G. York, J.~Adelman, Jr. Anderson, John~E., Scott~F. Anderson, James
  Annis, Neta~A. Bahcall, J.~A. Bakken, Robert Barkhouser, and et~al. Bastian,
  Steven.
\newblock The sloan digital sky survey: Technical summary.
\newblock {\em The Astronomical Journal}, 120(3):1579--1587, September 2000.

\bibitem{An2008}
Deokkeun An, Jennifer~A. Johnson, James~L. Clem, Brian Yanny, Constance~M.
  Rockosi, Heather~L. Morrison, Paul Harding, James~E. Gunn, and et~al.
  Allende~Prieto.
\newblock Galactic globular and open clusters in the sloan digital sky survey.
  i. crowded-field photometry and cluster fiducial sequences in ugriz.
\newblock {\em The Astrophysical Journal Supplement Series}, 179(2):326--354,
  December 2008.

\bibitem{atkinson2008introduction}
Kendall~E Atkinson.
\newblock {\em An introduction to numerical analysis}.
\newblock John wiley \& sons, 2008.

\bibitem{ascher1998computer}
Uri~M Ascher and Linda~R Petzold.
\newblock {\em Computer methods for ordinary differential equations and
  differential-algebraic equations}, volume~61.
\newblock Siam, 1998.

\bibitem{prothero_2016}
Richard Prothero.
\newblock Towns and cities analysis, england and wales, march 2016, Mar 2016.

\bibitem{ministry_of_housing_2019}
Communities and; Local~Government Ministry~of Housing.
\newblock English indices of deprivation 2019, Sep 2019.

\end{thebibliography}



\appendix

\section{Questionnaire}

Below are the questions used for the ThaiPASS'19 pre- and post-school surveys.

\noindent 1. Who are you? Please tick. \\
\indent A student    \\
\indent A teacher   \\
\indent NARIT staff \\                    
\indent Others (please specify) \\
 
\noindent 2. Please rate the following activities on a scale of 1 to 5, where: \\
 
\indent 5 = very interesting       \\
\indent 4 = interesting         \\
\indent 3 = ok         \\
\indent 2 = uninteresting  \\      
\indent 1 = very uninteresting \\

\noindent Please give comments if possible. e.g. what did you find interesting or not so interesting? Leave blank if you are unsure.
 
a)    Day 1: Why is Python useful + revision \\
\indent Rating:      1   \ \   2   \ \    3    \ \   4  \ \     5     \ \     (please circle) \\
\indent Comments: \\

b)    Day 2: Nuclear astrophysics \\
\indent Rating:      1   \ \   2   \ \    3    \ \   4  \ \     5     \ \     (please circle) \\
\indent Comments: \\

c)     Day 3: Half-day Python activities \\
\indent Rating:      1   \ \   2   \ \    3    \ \   4  \ \     5     \ \     (please circle) \\
\indent Comments: \\

d)    Day 4: Supernovae/Galaxies/Dark matter \\
\indent Rating:      1   \ \   2   \ \    3    \ \   4  \ \     5     \ \     (please circle) \\
\indent Comments: \\

e)    Day 5: Half-day “University and career” talks \\
\indent Rating:      1   \ \   2   \ \    3    \ \   4  \ \     5     \ \     (please circle) \\
\indent Comments: \\
 
The following questions were only used in the post-school survey. \\

3.  Did you learn any new skills / knowledge while at ThaiPASS? 
If YES, choose which skills. You can choose more than one answer. \\
\indent New programming skills \\
\indent New knowledge about astronomy \\
\indent New knowledge about data science \\
\indent Improved English communication skills \\
\indent Improved teamwork and collaboration skills \\
\indent I did not gain any new skills/knowledge \\
\indent Other (please describe) \\

If NO, please explain why. \\

4. Please choose one option per question. (1 = Strongly disagree, 3 = neither agree nor disagree, 5 = strongly agree). Leave blank if you are unsure.

a) ThaiPASS has improved my knowledge of Python. \\
\indent Rating:      1   \ \   2   \ \    3    \ \   4  \ \     5  \\

b) ThaiPASS has improved my knowledge on astronomy. \\
\indent Rating:      1   \ \   2   \ \    3    \ \   4  \ \     5  \\

c) ThaiPASS has increased my understanding of why data science is important for astronomy and other STEM subjects. \\
\indent Rating:      1   \ \   2   \ \    3    \ \   4  \ \     5  \\

c) I enjoyed the lectures at ThaiPASS. \\
\indent Rating:      1   \ \   2   \ \    3    \ \   4  \ \     5  \\

d) I enjoyed the Python coding sessions. \\
\indent Rating:      1   \ \   2   \ \    3    \ \   4  \ \     5  \\

e) I have a better understanding of how to use Python to plot graphs and present scientific data. \\
\indent Rating:      1   \ \   2   \ \    3    \ \   4  \ \     5  \\

f) ThaiPASS has encouraged me to study a STEM subject at university. \\
\indent Rating:      1   \ \   2   \ \    3    \ \   4  \ \     5  \\

g) ThaiPASS has made me interested in studying in the UK. \\
\indent Rating:      1   \ \   2   \ \    3    \ \   4  \ \     5  \\

5. What was the most enjoyable thing about ThaiPASS? \\

6. What could be improved for the next ThaiPASS? \\

\end{document}